\documentclass[twocolumn,showpacs,preprintnumbers,amsmath,amssymb,prb,aps]{revtex4-1}
\usepackage{epsfig}% Include figure files
\usepackage{graphicx}% Include figure files
\usepackage{dcolumn}% Align table columns on decimal point
\usepackage{bm}% bold math
\usepackage{textcomp}
\usepackage{subfig}
\begin{document}

\title{Understanding the Thermoelectric Properties of LaCoO$_{3}$ Compound}

\author{Saurabh Singh}
\altaffiliation{Electronic mail: saurabhsingh950@gmail.com}
\author{Sudhir K. Pandey}
\affiliation{School of Engineering, Indian Institute of Technology Mandi, Kamand - 175005, India}

\date{\today}

\begin{abstract}
  We present the thermoelectric (TE) properties of LaCoO$_{3}$ compound in the temperature range 300-600 K. The experimental value of Seebeck coefficient ($\alpha$) at 300 K is found to be $\sim$635 $\mu$V/K. The value of $\alpha$ decreases continuously with increase in temperature and reaches to $\sim$46 $\mu$V/K at $\sim$600 K. The electronic and TE properties of the compound have also been investigated by combining the \textit{ab-initio} electronic structures and Boltzmann transport calculations. LSDA plus Hubbard U (U= 2.75 eV) calculation on low spin configuration of the compound gives an energy gap of $\sim$0.5 eV, which is close to the experimentally reported energy gap. The effective mass of holes (\textit{m$^{*}_h$}) at $\Gamma$ point is nearly two times larger than the value of effective mass of electrons (\textit{m$^{*}_e$}) at FB point along the L and T directions, whereas the \textit{m$^{*}_e$} at FB point along the $\Gamma$ direction is nearly eight times larger than the value of \textit{m$^{*}_h$} at $\Gamma$ point along the FB direction. The large effective mass at FB point along the $\Gamma$ direction suggests that the TE property of this compound is mainly decided by the effective mass of the charge carriers in this direction. The calculated temperature dependent values of $\alpha$ are in fairly good agreement with experimental data in the temperature range 300-360 K, and above this temperature slight deviation is observed. The value of power factor (PF) for \textit{n}-type is $\sim$1.3 times larger the value of \textit{p}-type doped compound at 1100 K. The value of \textit{figure-of-merit} (\textit{ZT}) for \textit{n}-type doped compound is obtained $\sim$0.35 in the temperature range 600-1100 K, which suggests that with appropriate \textit{n}-type doping this compound can be used as a good TE material in the high temperature region.   

Keywords: Seebeck coefficient, Electronic structures, Thermoelectric properties, Oxide thermoelectric
\end{abstract}

%\pacs{71.20.-b, 71.15.Mb, 74.25.Fy}

\maketitle

\section{Introduction} 
  In the last few decades, fast consumption rate of available natural sources of energy (oils and gases) due to large need of mankind demands an alternate source of energy without affecting the environmental conditions \cite{DiSalvo99}. The various devices such as solar, hydro-power, bio-mass and thermoelectrc (TE) generators are being used to generate the required electricity. Among all these, TE generators have attracted much attention due to their potential capabilities of direct conversion of waste heat into useful electricity.$^{2,3}$ Using TE generators one can also utilize the waste heat generated from the other power generating sources and convert it into the electricity.$^{4,5}$ TE devices have various applications in electricity generation, TE cooler, temperature sensors etc.$^{6}$ There are various TE materials such as, lead and bismuth chalcogenides, silicon-germanium based materials, Clathrates, Half-Heusler alloys and organic TE materials have been explored for making the TE devices.$^{7}$ These materials have the limitations of use in high temperature region due to the chemical and thermal stabilities, high cost factor in the device fabrication, environmentally unfavourable and hazardous issues. To overcome these issues and have wide temperature TE applications in the high temperature region, the metal oxides give a new hope for the researchers.$^{8-10}$ In 1997, the discovery of \textit{p}-type Na$_{\textit{x}}$CoO$_{2}$ with a \textit{ZT} value $\sim$1 gave a breakthrough to explore more oxide materials for the TE applications.$^{11}$\\
 In high temperature region, oxide materials are more suitable for the TE applications due to their structural and chemical stabilities, oxidation resistance and low cost.$^{12-16}$ Therefore, it is important to estimate the physical parameters by which one can check the potential capabilities of oxide materials to be used for the TE applications. Any material is suitable for the TE applications is decided by the dimensionless parameter called as \textit{ZT}. The \textit{ZT} is defined as,$^{17,18}$\\
 \begin{equation}
 ZT= \alpha^{2}\sigma T/\kappa 
 \end{equation}
  where $\alpha$ is Seebeck coefficient (also known as thermopower), $\sigma$ and $\kappa$ ($\kappa$$_{l}$ +$\kappa$$e$) are the electrical and thermal conductivities, respectively. The term $\alpha$$^{2}$$\sigma$ in the numerator is known as power factor. For large value of \textit{ZT}, the material should have large PF and small $\kappa$.$^{19}$ The materials having large value of \textit{ZT} are more efficient for converting the waste heat into useful electricity.$^{20}$\\
  Transition metal oxides having Perovskite-based structure, have been extensively studied due to their flexible crystal structure and shows interesting physical properties such as colossal magnetoresistance, ferroelectricity, and thermoelectricity.$^{21,22}$ In the search of suitable oxides for TE applications, cobalt oxides having perovskite structure were found more interesting due to its large Seebeck coefficient and semiconducting or metallic electrical conductivity.$^{23-25}$ Lanthanum cobalt oxides, LaCoO$_{3}$ with rhombohedra distorted perovskite structure exhibits a large positive Seebeck coefficient at room temperature, with $\alpha$ = $\sim$640 μV/K.$^{26,27}$ The experimental and theoretical studies of electronic properties for this compound have been studied by many groups.$^{25-34}$ Although the thermopower, electrical and thermal conductivities of LaCoO$_{3}$ have been studied in the wide temperature range. However, a detailed understanding of the large and positive $\alpha$ observed in the compound by using the electronic structure calculation is still missing. This gives us a direction to investigate the high temperature TE properties of the LaCoO$_{3}$ by using experimental and theoretical tools.\\
  In the present work, we have performed the temperature dependent Seebeck coefficient ($\alpha$) measurement on LaCoO$_{3}$ compound in the range 300-600 K. The observed values of $\alpha$ at 300 and 600 K are $\sim$635 and $\sim$46 $\mu$V/K, respectively. The temperature dependent behavior of $\alpha$ is non-linear in the range 300-540 K, whereas it is almost linear in the range 540-600 K. To understand the TE behavior of this material, we have studied the electronic and TE properties by combining the \textit{ab-initio} electronic structures and Boltzmann transport calculations. The value of energy gap is found to be $\sim$0.5 eV, which is close to experimental band gap of $\sim$0.6 eV.$^{34,35}$ To know the contributions of charge carriers in $\alpha$, the total and partial density of states have been plotted for Co \textit{3d} orbitals and O \textit{2p} orbitals. The electronic band structure along high symmetric points is also plotted and effective masses of holes and electrons have been estimated from this. The values of $\alpha$ in the temperature range 300-1100 K are obtained from the BoltzTraP calculations. The calculated values of $\alpha$ show similar temperature dependent behavior as reported in literature. In the temperature range 300-360 K, the calculated values of $\alpha$ have good matching with experimental values. To see the effect of doping on TE behavior of the compound, we have plotted the chemical potential ($\mu$) dependent power factor (PF) at various temperatures. At 1110 K, the PF's value for positive $\mu$ comes out $\sim$1.3 times larger than the value of PF for negative $\mu$. The value of \textit{ZT} are evaluated in the temperature range 300-1100 K by using the peak value of power factor ($\alpha^{2}$$\sigma$) corresponding to positive value of chemical potential and thermal conductivities values reported by Pillai et al.$^{49}$ The value of \textit{ZT} is found to be $\sim$0.35 in the temperature range 600-1100 K.

  \section{Experimental and Computational details }
  
  The Polycrystalline LaCoO$_{3}$ sample was prepared by using the single step solution combustion method.$^{36}$ The detailed synthesis procedure of LaCoO$_{3}$ compound is given in our earlier work.$^{37}$ The sample used for $\alpha$ measurement was made in the pellet form under the pressure of $\sim$35 Kg/cm$^{2}$. Further, pellet was sintered at 1000 $^{0}$C for 24 hours. The diameter and thickness of the sample were $\sim$5 mm and $\sim$0.5 mm, respectively. The temperature dependent values of $\alpha$ were obtained in the temperature range 300-600 K by using the home-made Seebeck coefficient measurement setup.$^{37}$ \\
  The electronic structures calculations on LaCoO$_{3}$ compound in low spin configurations of Co$^{3+}$ ions (with \textit{$\text{t}^6_{2g}$}$\text{e}^0_{g}$$\Rightarrow$ S=0) were performed by using the full-potential linearized augmented plane-wave (FP-LAPW) method within the density functional theory (DFT) implemented in WIEN2k code.$^{38}$ The TE properties of the compound have been studied by using the BoltzTraP code.$^{39}$ LSDA+U approximation have been used to account the electronic correlations, where U is the on-site Coulomb interaction strength among Co \textit{3d} electrons.$^{40}$ The calculated energy band gap is found to be closer to the experimental value by considering the value of U = 2.75 eV.$^{35}$ The experimental lattice parameters corresponding to 300 K temperature and atomic coordinates of La, Co and O atoms used for the calculations were taken from the literature.$^{41}$ The muffin-tin sphere radii used in the calculations were set to 2.45, 1.94 and 1.67 Bohr for La, Co, and O atoms, respectively. The value of R$_{MT}$K$_{max}$, which determines the matrix size for convergence was set equal to 7, where R$_{MT}$ is the smallest atomic sphere radius and K$_{max}$ is the plane wave cut-off. The self-consistency was achieved by demanding the convergence of the total charge/cell to be less than 10$^{-4}$ electronic charge, which gives rise the energy convergence less than the 10$^{-5}$ Ry. The k-point mesh size was set to be 40$\times$40$\times$40 during the calculations of electronic and transport properties. The value of \textit{lpfac} parameter, which represents the number of k-points per lattice point, was set to be 5 for the calculation of TE properties.    
\section{Results and Discussion}
The temperature dependent $\alpha$ for the compound is shown in Fig. 1. 
\begin{figure}[htbp]
  \begin{center}
    \vspace{1.0cm}
   \includegraphics[width=0.40\textwidth]{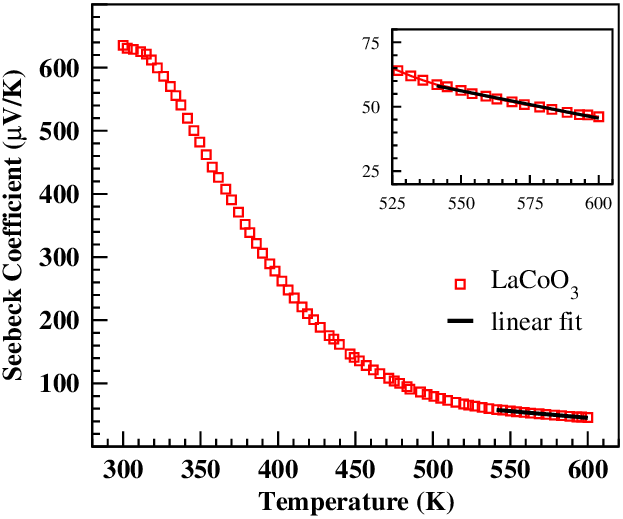}
    \label{}
    \captionsetup{justification=raggedright,
singlelinecheck=false
}
    \caption{(Color online) Temperature dependent Seebeck coefficient of LaCoO$_{3}$ compound.}
    \vspace{0.0cm}
  \end{center}
\end{figure} 
The values of $\alpha$ are positive in the 300-600 K temperature range. At 300 K, the observed value of $\alpha$ is $\sim$635 $\mu$V/K. As the temperature increases from 300 K to 600 K, the value of $\alpha$ decreases continuously. The observed value of $\alpha$ at 600 K is equal to $\sim$46$\mu$V/K. The temperature dependence variation in $\alpha$ is almost similar to the reported data in the literature.$^{26,27}$ The value $\alpha$ decreases with very fast rate in the temperature range 300-540 K and change of $\sim$576 $\mu$V/K is observed in the same temperature interval. For the temperature range 540-600 K, the change in the value of $\alpha$ is $\sim$13 $\mu$V/K. The variation in the $\alpha$ is non-linear in the temperature range 300-540 K. For temperature above 540 K, the values of $\alpha$ show almost a linear behavior. For the sake of clarity, the linear fit of experimental data is shown in the inset of Fig. 1. The linear temperature behavior above 540 K is associated with the insulator-to-metal transitions.$^{42}$
  
To understand the large and positive values of $\alpha$ exhibited by the compound, we have carried out the electronic structure calculations. The total density of states (TDOS) for the compound is shown in the Fig. 2a.
\begin{figure}[htbp]
  \begin{center}
   \includegraphics[width=0.35\textwidth, totalheight=0.45\textheight]{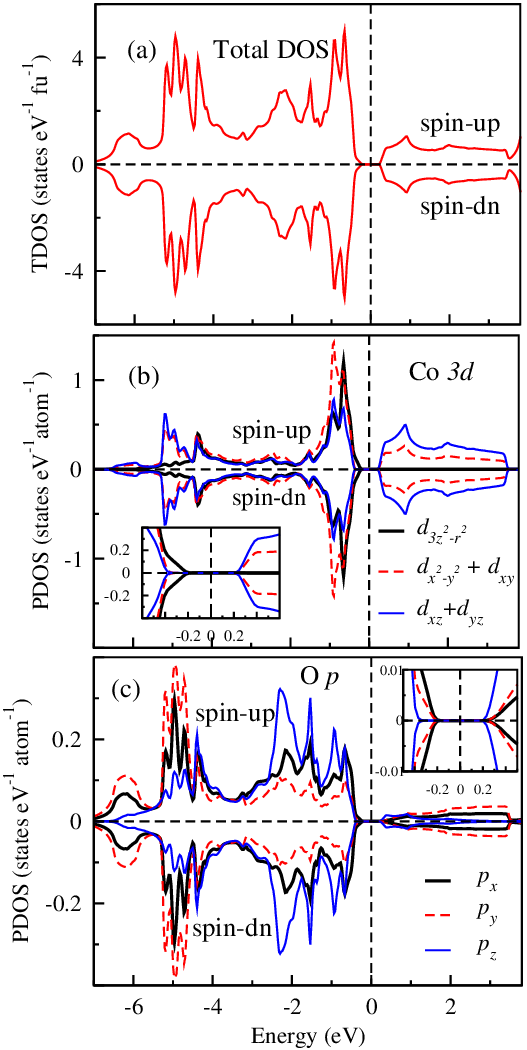}
    \label{}
    \captionsetup{justification=raggedright,
singlelinecheck=false
}
    \caption{(Color online) Total and partial density of states plots for LaCoO$_{3}$. Shown are (a) the TDOS plot, (b) PDOS of Co atom (\textit{3d} orbitals), (c) PDOS of O atom (\textit{2p} orbitals).}
    \vspace{0.0cm}
  \end{center}
\end{figure}
The electronic properties are calculated in the ground state (T= 0 K). Thus, the chemical potential in DOS plots can be taken at the middle of the energy gap. The dashed line represents the $\mu$, which is at the middle of energy gap. It is evident from the TDOS plot that LaCoO$_{3}$ is an insulating material with energy gap of $\sim$0.5 eV, which is in accordance to the observed energy gap.$^{35}$ In the low spin state, calculated DOS for up and down spins are found equal and opposite. Thus, it is evident from the TDOS plot that system has non-magnetic ground state.\\ 
In order to see the contributions of electrons to $\alpha$ from different atomic orbitals, we have also shown the partial density of states (PDOS) plots for Co and O atoms. The PDOS plot for Co \textit{3d} orbitals is shown in the Fig. 2b. For the rhombohedral lattice, five degenerate \textit{d} orbitals split into non-degenerate \textit{d$_{3z^{2}-r^{2}}$} orbital; doubly degenerate \textit{d$_{x^{2}-y^{2}}$} and \textit{d$_{xy}$} orbitals; and doubly degenerate \textit{d$_{xz}$} and \textit{d$_{yz}$} orbitals. Here, doubly degenerate \textit{d$_{x^{2}-y^{2}}$} and \textit{d$_{xy}$} orbitals are represented by \textit{d$_{x^{2}-y^{2}}$+d$_{xy}$}; and doubly degenerate \textit{d$_{xz}$} and \textit{d$_{yz}$} orbitals by \textit{d$_{xz}$}+\textit{d$_{yz}$} notation. From the PDOS of Co \textit{3d} orbitals, it is clearly observed that in valence band (VB) contributions in DOS are from all three states, \textit{d$_{3z^{2}-r^{2}}$}, \textit{d$_{x^{2}-y^{2}}$}+\textit{d$_{xy}$} and \textit{d$_{xz}$}+\textit{d$_{yz}$}, whereas in conduction band (CB) only \textit{d$_{x^{2}-y^{2}}$}+\textit{d$_{xy}$} and \textit{d$_{xz}$}+\textit{d$_{yz}$} states closer to Fermi level (E$_{F}$) contribute in the DOS. For the sake of clarity, the regions around top of the VB and bottom of the CB are shown in the inset of the Fig. 2b. At the top of VB, the contributions in DOS is only from the \textit{d$_{3z^{2}-r^{2}}$} orbital. Therefore, thermally excited electrons from this \textit{d$_{3z^{2}-r^{2}}$} orbital will contribute maximum to the value of $\alpha$. The energy of \textit{d$_{x^{2}-y^{2}}$}+\textit{d$_{xy}$} and \textit{d$_{xz}$}+\textit{d$_{yz}$ } states is $\sim$145 meV lower than that of \textit{d$_{3z^{2}-r^{2}}$} state. Thus, at 300 K thermally excited electrons from VB to the CB will be mainly from the \textit{d$_{3z^{2}-r^{2}}$} state. As the temperature increases, the number of thermally excited electrons from the \textit{d$_{x^{2}-y^{2}}$}+\textit{d$_{xy}$} and \textit{d$_{xz}$}+\textit{d$_{yz}$ } states will also be increase. The temperature corresponding to the energy gap of 145 meV is $\sim$1740 K. At this temperature, the transitions of thermally excited electrons from VB to CB will be almost equal in number. Thus at 1740 K, the contributions to $\alpha$ by thermally excited electrons will be nearly equal from the \textit{d$_{3z^{2}-r^{2}}$}, \textit{d$_{x^{2}-y^{2}}$}+\textit{d$_{xy}$} and \textit{d$_{xz}$}+\textit{d$_{yz}$ } states. At the bottom of the CB, there are two orbitals \textit{d$_{x^{2}-y^{2}}$}+\textit{d$_{xy}$} and \textit{d$_{xz}$}+\textit{d$_{yz}$}, which contribute to the DOS. Hence, these two orbitals will be mainly occupied by the thermally excited electrons from VB. There is no contribution in DOS from the \textit{d$_{3z^{2}-r^{2}}$} orbital at bottom of the CB. Thus, there will be no contribution to $\alpha$ by CB electrons from \textit{d$_{3z^{2}-r^{2}}$} orbital. At energy $\sim$80 meV from bottom of the CB, the DOS for \textit{d$_{x^{2}-y^{2}}$}+\textit{d$_{xy}$} orbitals have nearly two times larger value than \textit{d$_{xz}$}+\textit{d$_{yz}$} orbitals. Therefore, the  \textit{{d$_{x^{2}-y^{2}}$}+\textit{d$_{xy}$}} orbitals will occupy nearly twice the number of thermally excited electrons and thus have more contributions in the value of $\alpha$ LaCoO$_{3}$.\\
The PDOS of oxygen atom (\textit{2p} orbitals) is shown in the Fig. 2c. For both VB and CB, there are contributions in the DOS from all the three states of \textit{ p} orbital i.e. \textit{p$_{x}$}, \textit{p$_{y}$} and \textit{p$_{z}$}. The VB and CB regions about $\mu$ is shown in the inset of Fig. 2c for the sake of clarity. It is observed that the DOS contribution at top of the VB is from the \textit{p$_{x}$} and \textit{p$_{y}$} states, whereas at the bottom of CB \textit{p}$_{z}$ states contribute in the DOS. The DOS value of \textit{p$_{x}$} orbital near the top of VB are $\sim$10 times smaller than the DOS of \textit{d$_{3z^{2}-r^{2}}$} orbital. Therefore, the contributions in the value of $\alpha$ is very small from the \textit{p} orbital electrons. From the inset of Fig. 2b and 2c, the shape of DOS for different orbitals in VB clearly depicts the hybridization between \textit{p} and \textit{d} orbitals. In VB region closer to E$_{F}$, the \textit{d$_{3z^{2}-r^{2}}$} orbital hybridized with \textit{p$_{x}$} and \textit{p$_{y}$} orbitals, whereas \textit{d$_{x^{2}-y^{2}}$}+\textit{d$_{xy}$} and \textit{d$_{xz}$}+\textit{d$_{yz}$} orbitals hybridized with \textit{p}$_{z}$ orbital. In CB shape of PDOS closer to E$_{F}$ show that \textit{d$_{x^{2}-y^{2}}$}+\textit{d$_{xy}$} and \textit{d$_{xz}$}+\textit{d$_{yz}$} orbitals are hybridized with \textit{p$_{z}$} orbital.\\
The magnitude of $\alpha$ is decided by the effective mass, types of charge carriers and carriers concentrations, whereas its sign is decided by the effective mass of the charge carrier. To understand the large and positive values of $\alpha$ shown by the LaCoO$_{3}$, it is essential to calculate the effective masses of the holes and electrons. The effective masses of charge carriers can be estimated from the electronic band structure, therefore the dispersion curves for different k-values are calculated along the high symmetric points. The dispersion curves along the high symmetric $\Gamma$, T, L, and FB points are shown in Fig. 3. 
\begin{figure}[htbp]
  \begin{center}
  \vspace{1.0cm}
   \includegraphics[width=0.40\textwidth]{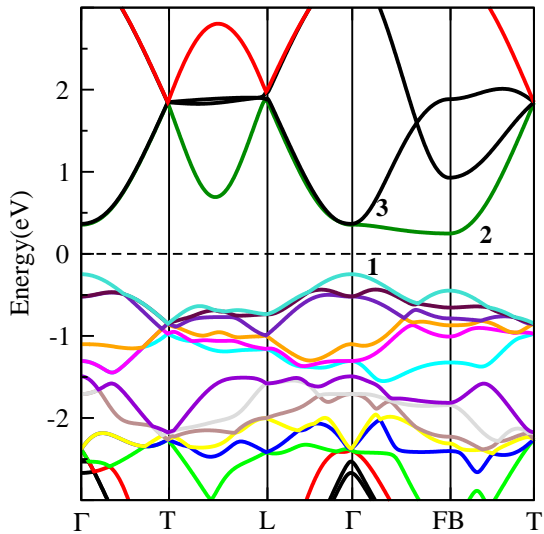}
    \label{}
    \captionsetup{justification=raggedright,
singlelinecheck=false
}
    \caption{(Color online) Dispersion curves along high symmetric points ($\Gamma$, T, L, and FB) for LaCoO$_{3}$ compound.}
    \vspace{0.1cm}
  \end{center}
\end{figure}
The primitive basis co-ordinates of these high symmetric points corresponds to R$\bar{3}$c space group are taken from the  Bilbao crystallographic server.$^{43}$It is evident from the dispersion curves, the top of the VB and bottom of the CB lies at $\Gamma$ and FB point, respectively. Therefore, LaCoO$_{3}$ compound is an indirect band gap semiconductor. The estimated energy of indirect band gap from the dispersion curve is found to be 0.5 eV. The top of the VB (Band \textit{\textbf{1}}) and bottom of the CB (Band \textit{\textbf{2}}) are non-degenerate. The minimum of second band (i.e. Band \textbf{\textbf{3}}) from the bottom of CB lies at the $\Gamma$ point. The energy gap between top of the VB and bottom of the CB (Band \textit{\textbf{3}}) is $\sim$ 0.6 eV. This energy gap is corresponding to the temperature of $\sim$7200 K. The band \textit{ \textbf{2}} and \textit{ \textbf{3}} in the CB are doubly degenerate at $\Gamma$ point. The magnitude of effective mass at a given k-point is decided by the shape of energy band. At a given k-point, the flat band energy curve has the larger effective mass in comparison to the narrower energy curve. From the dispersion curve, it is observed that at FB point the energy curve (Band \textit{\textbf{2}}) have more flat shape along the $\Gamma$ direction in comparison of the curve along T direction. Therefore, one expect a large effective mass of the electrons at FB point along the $\Gamma$ direction. For the VB energy curve (Band \textit{\textbf{1}}), the shape of the energy band at $\Gamma$ point is almost symmetric along FB and L directions. Thus, the effective mass of the holes will be nearly equal along these directions. The shape of band \textbf{\textit{1}} at $\Gamma$ point and along the FB direction is more flat than the shape of band \textit{\textbf{2}} at FB point along the T direction, whereas the shape of energy curve of band \textit{\textbf{1}} at $\Gamma$ point along L direction is narrower than the shape of band \textit{\textbf{2}} at FB point along $\Gamma$ direction. Thus, the effective mass of electrons at FB point in band \textit{\textbf{2}} and along $\Gamma$ direction is expected very large than that of holes effective mass at $\Gamma$ point along L direction.\\  
To understand the effective mass contributions to $\alpha$ from the charge carriers (holes and electrons), we have estimated the effective mass along the high symmetric direction by using the dispersion curve. The effective mass of holes and electrons are calculated at $\Gamma$ point (along L, FB, $\&$ T directions) and FB point (along $\Gamma$, L, $\&$ and T directions), respectively. The calculated effective mass is given in the Table I.
\vspace{0.2cm}
 \begin{table}[htbp]
\caption{The effective mass of holes (at $\Gamma$  point) and electrons (at FB point) along high symmetric directions.}   
\begin{tabular}{p{0.cm}p{2cm}p{0.5cm}p{0.2cm}p{0.2cm}p{0.5cm}}
\hline
\hline
&&\multicolumn{2}{c}{Effective mass (m$^{*}$/m$_{e}$)}\\
\cline{3-4}
\\
&High-Symmetry Point&&Valence Band&Conduction Band\\
&&{ {\it } }&&&\\
\hline
$\Gamma$$-$$\Gamma$\textbf{L}&&&1.87&-&\\
$\Gamma$$-$$\Gamma$\textbf{FB}&&&1.82&-&\\
$\Gamma$$-$$\Gamma$\textbf{T}&&&1.76&-&\\
FB$-$FB$\boldsymbol{\Gamma}$&&&-&14.70&\\
FB$-$FB\textbf{L}&&&-&0.84&\\
FB$-$FB\textbf{T}&&&-&1.01&\\

\hline
\hline
\end{tabular}
\end{table}
The notation $\Gamma$$-$$\Gamma$\textbf{L} used in the table represents the effective mass calculated at $\Gamma$ point along the L direction and similarly for others. The calculated effective mass of holes at $\Gamma$ point along L and T direction are nearly two times larger of the electron's effective mass at FB point, whereas effective mass of electrons at FB point along the $\Gamma$ direction is nearly eight times larger of the effective mass of holes at $\Gamma$ point along the FB direction. From the estimated values of effective masses of holes and electrons along different directions, it is expected that electrons will be the dominating charge carriers in the contributions to the thermopower due to their large effective mass (m$^{*}_e$ = $\sim$14.7m$_{e}$) at FB point along the $\Gamma$ direction. Thus, one expect large and negative $\alpha$ for the LaCoO$_{3}$ compound. From the calculations, the large and negative value of $\alpha$ is also obtained at middle of the energy band gap (i.e. $\mu$= 0 eV). But, the experimentally observed value of $\alpha$ for LaCoO$_{3}$ is large and positive. At this point one should keep in mind that $\mu$ for an intrinsic semiconductor materials has temperature dependence through the following expression$^{44}$\\
\begin{equation}
  \mu = \varepsilon_{v} + \frac{1}{2}Eg + \frac{3}{4}\textit{k}_{B}T \textit{ln} ({\textit{m}_{\textit{v}}}/{\textit{m}_{\textit{c}}})
  \end{equation}
 where $\varepsilon$$_{v}$ is the energy of the electron at the top of the VB, ${m_{v}}$ and ${m_{c}}$ are the effective mass of the charge carrier in VB and CB, respectively. Using the effective mass of the electrons, \textit{m$^{*}_e$} = 14.7 \textit{m$_{e}$}, and effective mass of the holes, \textit{m$^{*}_h$} = 1.8 \textit{m$_{e}$}, the value of $\mu$ at 300 K comes out to be $\sim$ -40 meV. The shift of $\mu$ below the middle of the gap represents the hole type behavior. Thus, one should expect a positive value of $\alpha$ for the LaCoO$_{3}$. The variations of $\alpha$ with $\mu$ at various temperatures is shown in the Fig. 4.
\begin{figure}[htbp]
  \begin{center}
   \vspace{1.5cm}
   \includegraphics[width=0.45\textwidth]{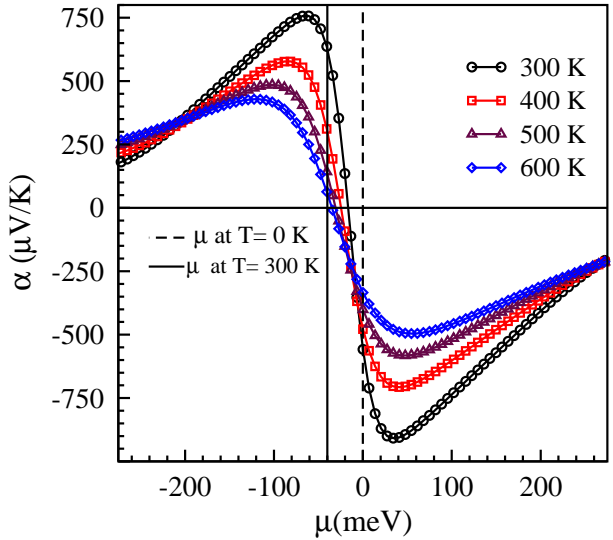}
    \label{}
    \captionsetup{justification=raggedright,
singlelinecheck=false
}
    \caption{(Color online) Variations of Seebeck coefficient with chemical potential for LaCoO$_{3}$ compound at different temperatures (300, 400, 500, and 600 K).}
    \vspace{0.0cm}
  \end{center}
\end{figure}

  \begin{figure}[htbp]
  \begin{center}
    
   \includegraphics[width=0.45\textwidth]{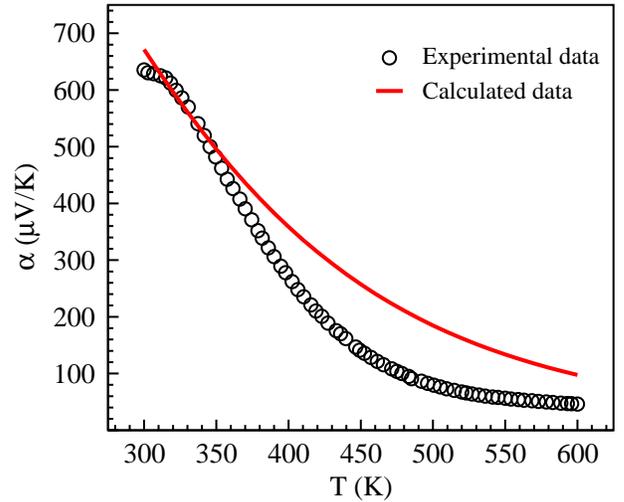}
    \label{}
    \captionsetup{justification=raggedright,
singlelinecheck=false
}
    \caption{(Color online) Seebeck coefficient ($\alpha$ vs temperature for LaCoO$_{3}$ compound. Shown Experimental data(O), and calculated data (\textbf{\textemdash}).}
    \vspace{0.0cm}
  \end{center}
\end{figure}
  The vertical solid line drawn at -40 meV represents the chemical potential at 300 K. The calculated value of $\alpha$ for $\mu$ = -40 meV is found to be $\sim$670 $\mu$V/K, which is closer to the experimental value $\sim$635$\mu$V/K. The calculated $\alpha$'s value in the temperature range 300-600 K along with experimental data for a fixed $\mu$ ( i.e. $\mu$ = -40 meV from the middle of the energy gap) are shown in the Fig. 5. The values of $\alpha$ obtained from the calculations have reasonably good match with experimental data in the temperature range 300-360 K. Above 360 K, difference in calculated and experimental data is observed. In our earlier work, we observed a better matching between calculated and experimental data for ZnV$_{2}$O$_{4}$ compound in the temperature range 200-400 K.$^{45}$ From the Fig. 5, we observed the difference of $\sim$120 $\mu$V/K for LaCoO$_{3}$ at temperature $\sim$460 K, whereas in the metallic phase, $\sim$600 K, this difference decreases to the value of $\sim$50 $\mu$V/K. In the present case, differences observed between calculated and experimental values of $\alpha$ above 360 K can be due to the various factor, such as relaxation time, spin state transition and temperature independent $\mu$. In the high temperature region, the relaxation time changes due to the various scattering mechanism.$^{46,47}$ Therefore, corrections in the relaxation time due to the different scattering mechanism is required. In this compound, the spin state transitions from low-spin to high-spin states are also reported in the temperature region 300-600 K.$^{25,48}$ In present case, we have performed the theoretical calculations for low spin configuration only. Hence, electronic structures calculations for different spin states in accordance to the temperature range may improve the data. Here to make the comparison between calculated and experimental data, we have shown $\alpha$ vs T plot for a fixed $\mu$. But, the $\mu$ for an intrinsic semiconductor is temperature dependent.$^{44}$ Therefore, for temperature dependent transport properties one should consider the temperature dependent correction in the $\mu$ also. In high temperature region, consideration of these three factors may lead to the better matching between calculated and experimental data. To established this conjecture careful study in this directions is needed, which is beyond the scope of the present work.\\
  \begin{figure}[htbp]
  \begin{center}
  
   \includegraphics[width=0.45\textwidth]{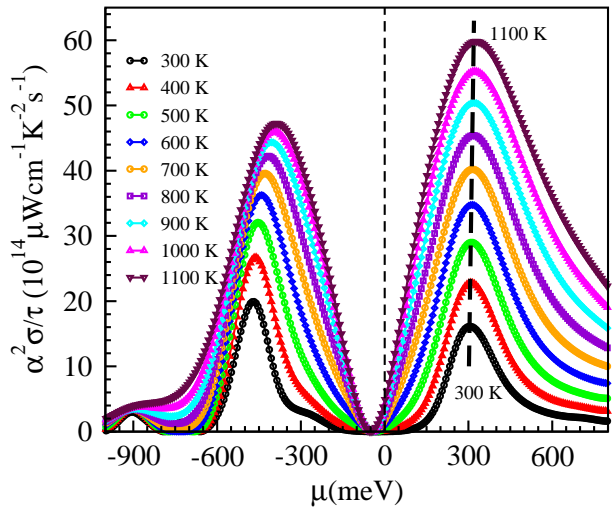}
    \label{}
    \captionsetup{justification=raggedright,
singlelinecheck=false
}
    \caption{(Color online) The variations in power factor ($\alpha$$^{2}$$\sigma$/$\tau$) with chemical potential at different temperatures.}
    \vspace{0.0cm}
  \end{center}
\end{figure}
In order to see the potential capabilities of the LaCoO$_{3}$ compound to be used for TE applications, we have also calculated the power factor (PF). Fig. 6 shows the variations in PF, $\alpha$$^{2}$$\sigma$/$\tau$, with $\mu$ at different temperatures.\\
  For each curve plotted at different temperatures, the value of PF have maximum for both positive and negative value of $\mu$. The positive and negative value of $\mu$ represents the electrons and holes type doping, respectively. The dotted line at $\mu$ equal to zero eV represents the middle of the energy band gap. The value of PF maximum for negative $\mu$ is found to be $\sim$19$\times$10$^{-14}$ $\mu$W cm$^{-1}$ K$^{-2}$ s$^{-1}$ at $\sim$ -467 meV for 300 K. As the temperature increases from 300 to 1100 K, the values of $\mu$ corresponds to maximum value of PF shifted from $\sim$ -467 meV to $\sim$ -386 meV, and the value of PF maximum at 1100 K is $\sim$47 $\times$10$^{-14}$ $\mu$W cm$^{-1}$ K$^{-2}$ s$^{-1}$. For the positive $\mu$, the maximum value for 300 K temperature curve is found at $\sim$307 meV, and it shifted towards higher value of $\mu$ as the temperature increases from 300 K to 1100 K. The maximum values of PF for 300 and 1100 K temperature curve are $\sim$16$\times$10$^{-14}$ and $\sim$60$\times$10$^{-14}$ $\mu$W cm$^{-1}$ K$^{-2}$ s$^{-1}$, obtained at the positive $\mu$ of $\sim$307 and $\sim$328 meV, respectively. The calculated values of PF suggests that an appropriate amount of \textit{n}-type doping can increase the value of PF and hence the TE behavior of this compound can be enhanced.\\
We have also estimated the values of \textit{ZT} for the \textit{n}-type doped LaCoO$_{3}$ compound by using the maximum values of PF corresponds to positive $\mu$ (shown in Fig. 6) and thermal conductivity data of compound reported by Pillai et. al.$^{49}$ Here, the value of relaxation time ($\tau$) is taken equal to 10$^{-14}$ s. The values of \textit{ZT} have been estimated in the temperature range 300-1100 K at the temperature interval of 100 K. The $\kappa$ values used for calculation are obtained from the interpolation of $\kappa$. Temperature dependent \textit{ZT} of LaCoO$_{3}$ is shown in the Fig. 7. 
\begin{figure}[htbp]
  \begin{center}
     \vspace{1.0cm}
   \includegraphics[width=0.40\textwidth]{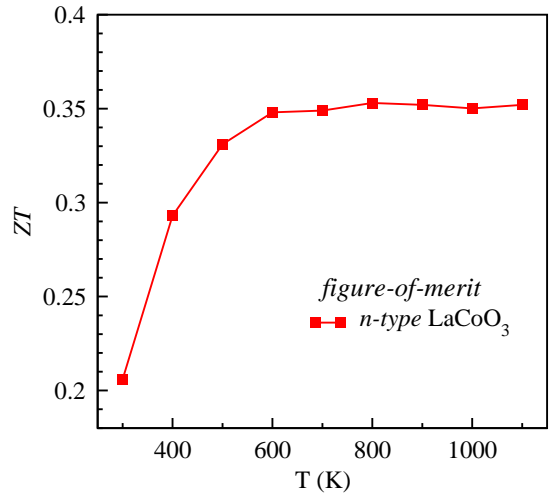}
    \label{}
    \captionsetup{justification=raggedright,
singlelinecheck=false
}
    \caption{(Color online) Temperature dependent variation of calculated \textit{figure-of-merit}, \textit{ZT}, for n-type LaCoO$_{3}$. The values of \textit{ZT} are obtained by using peak value of power factor corresponding to positive value of chemical potential(shown in Fig. 6), $\tau$ =10$^{-14}$ seconds, and thermal conductivities values reported by Pillai et al [Ref. \textbf{49}]}
    \vspace{0.0cm}
  \end{center}
\end{figure}
The value of \textit{ZT} at 300 K is $\sim$0.21 and it increases to $\sim$0.35 at 600 K. Further increase in the temperature above 600 K, ZT becomes almost constant upto 1100 K. The value of \textit{ZT} suggests that with suitable \textit{n}-type doping, LaCoO$_{3}$ compound can be used as a better TE material in high temperature region. The value of \textit{ZT} for \textit{n}-type doped LaCoO$_{3}$ in the temperature range 600-1100 K is similar to the other oxides, such as La/Nb doped SrTiO$_{3}$, Ge doped In$_{2}$O$_{3}$, Nb doped CaMnO$_{3}$, Ga doped Ca$_{3}$Co$_{4}$O$_{9}$.$^{12}$ At this point, it is important to notice that the values of \textit{ZT} are calculated by taking $\kappa$ values for the poly-crystalline sample.$^{49}$ By nano-structuring there are possibilities of reducing the $\kappa$ values.$^{50-53}$ Thus, \textit{ZT} of the \textit{n}-type doped LaCoO$_{3}$ can be further enhanced.
\section{Conclusions} 
In conclusion, we have studied the temperature dependent thermopower of LaCoO$_{3}$ compound in the temperature range 300-600 K. The values of $\alpha$ for the compound at 300 and 600 K are found to be $\sim$635 and $\sim$46 $\mu$V/K, respectively. For the temperature range 300-540 K, the temperature variations in the $\alpha$ is non-linear and the change of $\sim$576 $\mu$V/K is observed. In the 540-600 K, the change in the $\alpha$ is $\sim$13 $\mu$V/K and the temperature dependent variations is almost linear. The linear variations of $\alpha$ in the 540-600 K temperature range is due to insulator-to-metal transitions at $\sim$540 K. To understand the large and positive values of $\alpha$ shown by the LaCoO$_{3}$, we have also carried out the electronic structure and transport coefficients calculations by using the \textit{ab-initio} density functional theory and BoltZtraP Transport calculations, respectively. The estimated energy gap for U = 2.75 eV is equal to 0.5 eV, which is in accordance to the experimentally reported energy band gap. The flat energy curve observed in the band structure gives $\sim$8 times larger effective mass of electrons in conduction band than the effective mass of holes in valence band. The large effective mass of electron plays an important role in understanding the large and positive Seebeck coefficient exhibited by this compound. Further we have calculated the power factor, which suggests that the \textit{n}-type doping is more appropriate to enhance the TE behavior. We have also estimated the \textit{ZT} values for \textit{n}-type doped LaCoO$_{3}$. The obtained value of \textit{ZT} shows that, the compound can be used as a good TE material in the high temperature region 600-1100 K. Further, by nano-structuring one can reduce the thermal conductivity in high temperature region. Hence, LaCoO$_{3}$ compound can be a potential TE material for high temperature TE applications.

\end{document}